\def \SAIT #1 #2 {{\em Mem.\ Soc.\ Astron.\ It.\/} {\bf #1}, #2}
\def \MESS #1 #2 {{\em The Messenger\/} {\bf #1}, #2}
\def \ASTRNACH #1 #2 {{\em Astron. Nach.\/} {\bf #1}, #2}
\def \AAP #1 #2 {{\em Astron. Astrophys.\/} {\bf #1}, #2}
\def \AAL #1 #2 {{\em Astron. Astrophys. Lett.\/} {\bf #1}, L#2}
\def \AAR #1 #2 {{\em Astron. Astrophys. Rev.\/} {\bf #1}, #2}
\def \AAS #1 #2 {{\em Astron. Astrophys. Suppl. Ser.\/} {\bf #1}, #2}
\def \AJ #1 #2 {{\em Astron. J.\/} {\bf #1}, #2}
\def \ANNREV #1 #2 {{\em Ann. Rev. Astron. Astrophys.\/} {\bf #1}, #2}
\def \APJ #1 #2 {{\em Astrophys. J.\/} {\bf #1}, #2}
\def \APJL #1 #2 {{\em Astrophys. J. Lett.\/} {\bf #1}, L#2}
\def \APJS #1 #2 {{\em Astrophys. J. Suppl.\/} {\bf #1}, #2}
\def \APSS #1 #2 {{\em Astrophys. Space Sci.\/} {\bf #1}, #2}
\def \ASR #1 #2 {{\em Adv. Space Res.\/} {\bf #1}, #2}
\def \BAIC #1 #2 {{\em Bull. Astron. Inst. Czechosl.\/} {\bf #1}, #2}
\def \JSQRT #1 #2 {{\em J. Quant. Spectrosc. Radiat. Transfer\/} {\bf #1}, #2}
\def \MN #1 #2 {{\em Mon. Not. R. Astr. Soc.\/} {\bf #1}, #2}
\def \MEM #1 #2 {{\em Mem. R. Astr. Soc.\/} {\bf #1}, #2}
\def \PLR #1 #2 {{\em Phys. Lett. Rev.\/} {\bf #1}, #2}
\def \PASJ #1 #2 {{\em Publ. Astron. Soc. Japan\/} {\bf #1}, #2}
\def \PASP #1 #2 {{\em Publ. Astr. Soc. Pacific\/} {\bf #1}, #2}
\def \NAT #1 #2 {{\em Nature\/} {\bf #1}, #2}

\documentstyle[twoside]{memsait}
\input epsf.sty
\begin{opening}
\title{HIGHLY IONIZED NEON IN THE SPECTRUM OF THE SEY 2 GALAXY NGC 4507 ?}
\author{C. VIGNALI$^1$, R. AUDANO$^1$, A. COMASTRI$^2$, M. CAPPI$^3$, 
G. MATT$^{4}$}
\institute{$^1$Dipartimento di Astronomia, Universit\`a di Bologna, 
via Zamboni 33, I-40126 Bologna, Italy\\
$^2$Osservatorio Astronomico, via Zamboni 33, I-40126 Bologna, Italy\\
$^3$ITeSRE/CNR, via Gobetti 101, I-40129 Bologna, Italy\\
$^4$Universit\`a degli Studi ``ROMA TRE",  
via della Vasca Navale 84, I-00146 Roma, Italy}
\date{} 
\end{opening}

\begin{document}

\oddpagefooter{}{}{} 
\evenpagefooter{}{}{} 
\ 
\bigskip

\begin{abstract}
Results of an ASCA X-ray observation of NGC 4507 are presented. 
The spectrum is best parameterized by a double power law (or a partial 
covering) model plus a narrow FeK$\alpha$ emission line likely due to the 
line-of-sight absorption matter. The data require also an emission 
line at $\sim$ 0.9 keV consistent with NeIX, which may indicate that the soft 
X-ray emission derives from a combination of resonant scattering and 
fluorescence of the radiation by photoionized gas. 
Moreover, complex absorption (or other, unresolved, emission lines)
is required by the data below 3 keV. Similarities 
between the X-ray spectra of NGC 4507 and NGC 4151 are also stressed.
\end{abstract}

\section{NGC 4507: spectral analysis}
NGC 4507, a highly absorbed Sey 2 galaxy (z=0.012), was observed with 
the gas imaging spectrometer (GIS) and solid 
state spectrometer (SIS) on board the ASCA satellite (Tanaka et al. 1994) 
in February 1994. 
After applying standard selection criteria, 25 Ks for each SIS and 40 Ks for 
the GIS2 detector were collected.\\
Previous analisies 
showed a complex X-ray spectrum. Ginga observation revealed a hard power law 
($\Gamma_{\rm 2-20keV}$ $\sim$ 1.4$\pm{0.2}$), a high column density 
($N_{\rm H}$ $\sim$ 3.7$\pm{0.5}$ $\times$ 10$^{23}$ cm$^{-2}$) and an 
iron emission line (EW $\sim$ 400$\pm{100}$ eV) (Awaki et al. 1991), while 
OSSE observation (Bassani et al. 1995) showed a photon index $\Gamma$ 
$\sim$ 2.1$\pm{0.3}$ in agreement with those of Seyfert galaxies in the same 
energy range (Johnson et al. 1993). The ASCA data require at least 
a double power law (or partial covering) model plus a narrow FeK$\alpha$ line, 
whose rest energy indicates emission from neutral matter. The relevant data are 
listed in table 1. Moreover, from the spectrum below 1 keV there is evidence for
a soft narrow emission line at $\sim$ 0.9 keV (fig. 1). 
When it is fitted to the data, the improvement is significant 
($\Delta$$\chi^{2}$ $\sim$ 14 with the addition of 2 dof). 
The line energy is consistent with NeIX (E $\sim$ 0.92 keV, fig. 1) and may be 
produced in photoionized gas 
(i.e. which may be identified with the warm absorber observed in Sey 1 
galaxies) with contributions from both the resonant scattering and fluorescent
lines. The equivalent widths (EW) of the two components depends strongly 
on the optical depth of the emitting matter. The total observed value 
(i.e about
100 eV) suggests that the matter is completely thick to resonant absorption
and mildly thick to fluorescence, a situation which is realized if the
equivalent hydrogen column density of the matter 
is of the order of 10$^{22\div23}$ cm$^{-2}$, the exact value depending on 
geometrical and physical parameters. It is worth noting that a
similar estimate can be independently derived from the ratio of the scattered
and direct continua, which is about 1\%. 
 
The evidence for some residuals 
in the spectrum below 3 keV suggests complex 
absorption of the nuclear radiation, e.g. a dual absorber, with the second
absorbing matter possibly ionized. It is interesting to note that 
NGC 4507 shows strong similarities with NGC 4151 
(Leighly et al., in preparation). 
In both sources a ``soft" narrow line clearly 
improves the fit and the residuals indicate complex 
absorption below $\sim$ 3 keV. Soft emission lines have probably been 
detected by ASCA in other Seyferts like Mkn 3 (Iwasawa et al. 1994), NGC 4051 
(Guainazzi et al. in press) and NGC 4388 (Iwasawa et al. in press), but 
further investigations are needed to fully understand the nature of the 
soft X-ray emission.

\centerline{\bf Tab. 1 - Spectral parameters}

\begin{table}[h]
\hspace{1.5cm} 
\begin{tabular}{|l|c|c|c|c|c|c|c|}
\hline
\multicolumn{8}{|c|}{\bf NGC 4507} \\
\hline
model&$\Gamma_{1}$&$E_{\rm Ne}$&$EW_{\rm Ne}$&$N_{\rm H}$
&$\Gamma_{2}$&$E_{K\alpha}$&$EW_{K\alpha}$\\
\hline
po+po&2.34$^{+0.25}_{-0.20}$&\dotfill&\dotfill
&3.10$^{+0.37}_{-0.35}$$\times$10$^{23}$&1.80$^{+0.21}_{-0.22}$&6.36$^{+0.02}_{-0.03}$
&177$^{+47}_{-40}$\\
\hline
po+po+Ne&2.07$^{+0.24}_{-0.22}$&0.90$\pm{0.03}$&117$^{+49}_{-47}$&
''&''&''&''\\
\hline
\end{tabular}
\end{table}
%

%
%
\begin{figure}
\epsfysize=4.8cm 
\hspace{0.65cm}\epsfbox{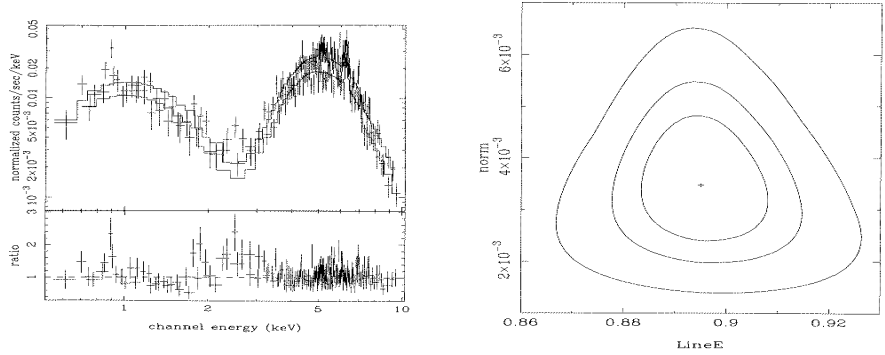}
\caption[h]{NGC 4507: ASCA spectrum; energy vs. intensity conf. cont. 
for the NeIX line.}
\end{figure}
%
%
\vspace{-0.5cm} 


\begin{thebibliography}{99}  

\bibitem[]{}Awaki, H., et al.: 1991, \PASJ 43 L37
\bibitem[]{}Bassani, L., et al.: 1995, \APJL 444 73
\bibitem[]{}Guainazzi, M., et al.: 1996, {\em Publ. Astron. Soc. Japan}, 
{\bf 48}, in press
\bibitem[]{}Iwasawa, K., et al.: 1994, \PASJ 46 L167
\bibitem[]{}Iwasawa, K., et al.: 1996, in press
\bibitem[]{}Johnson, W.N., et al.: 1993, {\em Bull. Amer. Astron. Soc.}, 
{\bf 183}, 6403 
\bibitem[]{}Tanaka, Y., Inoue, H., Holt, S.S.: 1994, \PASJ 46 L37

\end{thebibliography}
\end{document}